\documentclass[twocolumn,floats,prl,aps,showpacs]{revtex4}
\usepackage{graphicx}
\usepackage{dcolumn} 
\usepackage{amsmath}
\usepackage{exscale}
\usepackage{bm}      
\usepackage{color}
\usepackage{epsfig,psfrag,subfigure,subeqnarray}
\usepackage{bbm}
\usepackage{amsbsy}
\usepackage{amssymb}
\usepackage{enumitem}

\def\lan{\langle}
\def\ran{\rangle}
\def\va{\varepsilon}

\def\vk{{\bf k}}
\def\vK{{\bf K}}

\def\vr{{\bf r}}
\def\vu{{\bf u}}
\def\vQ{{\bf Q}}
\def\vq{{\bf q}}
\def\vp{{\bf p}}
\def\vR{{\bf R}}

\newcommand{\bd}{\begin{equation}}
\newcommand{\ed}{\end{equation}}
\newcommand{\be}{\begin{equation}}
\newcommand{\ee}{\end{equation}}
\newcommand{\bt}{\begin{split}}
\newcommand{\et}{\end{split}}

\newcommand{\bn}{\begin{align}}
\newcommand{\en}{\end{align}}

\newcommand{\bea}{\begin{eqnarray}}
\newcommand{\eea}{\end{eqnarray}}
\newcommand{\ba}{\begin{array}}
\newcommand{\ea}{\end{array}}
\newcommand{\nn}{\nonumber}

\begin{document}

\title{Way to observe the implausible ``trion-polariton"}
\author{ Shiue-Yuan Shiau$^{1}$}\email{shiau.sean@gmail.com} 
\author{Monique Combescot$^{2}$, Yia-Chung Chang$^{3,1}$}
\affiliation{(1) Department of Physics, National Cheng Kung University, Tainan, 701 Taiwan}
\affiliation{(2) Institut des NanoSciences de Paris, Universit\'e Pierre et Marie Curie, CNRS, 4 place Jussieu, 75005 Paris}
\affiliation{(3) Research Center for Applied Sciences, Academia Sinica, Taipei, 115 Taiwan}

\date{\today }

\begin{abstract}
Using the composite boson (coboson) many-body formalism, we determine under which conditions ``trion-polariton" can exist.
Dipolar attraction can bind an exciton and an electron into a trion having an energy well separated from the exciton energy. Yet, the existence of long-lived ``trion-polariton" is a priori implausible not only because the photon-trion coupling, which scales as the inverse of the sample volume, is vanishingly small, but mostly because this coupling is intrinsically ``weak". Here, we show that a moderately dense Fermi sea renders its observation possible: on the pro side, the Fermi sea overcomes the weak coupling by pinning the photon to its momentum through Pauli blocking; it also overcomes the dramatically poor photon-trion coupling by providing a volume-linear trion subspace to which the photon is coherently coupled. On the con side, the Fermi sea broadens the photon-trion resonance due to the fermionic nature of trions and electrons; it also weakens the trion binding by blocking electronic states relevant for trion formation. As a result, the proper way to observe this novel polariton is to use doped semiconductor having long-lived electronic states, highly-bound trion and Fermi energy as large as a fraction of the trion binding energy.
\end{abstract}


\maketitle
\date{\today }
Exciton-polaritons have recently attracted very much attention because of claimed observations of Bose-Einstein condensation\cite{polCond1,polCond2,polCond3,polCond4}.
When the coupling between a photon and an exciton is strong, exciton-polaritons\cite{polPredict,polSeen} are formed. By contrast, when this coupling is weak, photons are absorbed. In the former case, the exciton lifetime is long compared to the Rabi oscillation period and the \textbf{Q} exciton recombines into the same \textbf{Q} photon (see Fig.~1(a)). When it is short, the exciton changes momentum before recombination occurs and the initial photon \textbf{Q} cannot be re-emitted; it gets absorbed along the Fermi golden rule, the small exciton lifetime producing a broadening of the exciton discrete level, which plays the role of a continuum\cite{DubinEPL}.

\begin{figure}[t!]
\centering
  \includegraphics[trim=4cm 2.5cm 4cm 3cm,clip,width=3in] {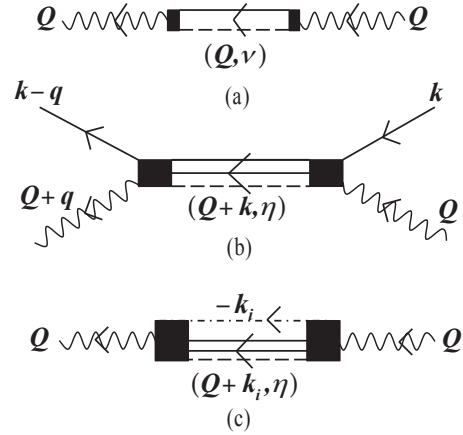}
 \vspace{-0.1cm}
\caption{\small (a) A $\vQ$ photon couples to excitons  having a center-of-mass momentum $\vQ$ and a relative-motion index $\nu$. The same $\vQ$ photon is emitted when the exciton lifetime is long on the Rabi-coupling scale. (b) A $\vQ$ photon and a $\vk$ electron couple to trions having a center-of-mass momentum $\vK=\vQ+\vk$ and a relative-motion index $\eta$. The emitted-photon momentum $\vQ+\vq$ can differ from $\vQ$  even when the trion lifetime is long for the trion to keep its $\vK$ momentum. (c): In the presence of a Fermi sea having $N$ electrons $\vk_i$, a $\vQ$ photon couples to any pair of $(\vQ+\vk_i,\eta)$ trion and $(-\vk_i)$ conduction hole.  }
\label{fig:1}
\end{figure}

We here consider another semiconductor bound state, the trion, and determine under which conditions this composite fermion can strongly couple to photon to form a polariton. Many objections lead us to first reject the idea: (i) the trion coupling to photon is intrinsically weak because the emitted photon can have a momentum different from its initial value, even when the trion lifetime is long; (ii) the photon-trion coupling is vanishingly small because it scales as the inverse of the sample volume; (iii) an increase of the number of electrons available for pairing broadens the photon-trion resonance due to the fermionic nature of trions and electrons; (iv) it also reduces the trion binding by Coulomb screening and by Pauli blocking the electronic states relevant to the formation of bound trion. Despite all these objections, we will show that there  exists a narrow window in which ``trion-polariton" can be formed. 

Semiconductor trions\cite{trionPredict,trionseen1,trionseen2,trionseen3,trionseen4,trionseen5,trionseen6} are made of two conduction-electrons and one valence-hole, or one conduction-electron and two valence-holes. We will here focus on the former case, denoted as
$X^-$.
Since orbital wave functions of ground states are even with respect to exchange, the two conduction-electrons of a $X^-$ ground-state trion must be in a spin-singlet state; they thus have opposite spins. While excitons result from the attraction of a conduction-electron by a valence-hole, trions result from the attraction of a conduction-electron by an excitonic dipole; this is why trion binding energies are usually small, typically one order of magnitude smaller than for exciton.  Yet, trion binding as large as 30 meV can be found in transition metal dichalcogenide materials\cite{Sidler2016,Jones2013,Ross2013,Mak2013,Xu2014}.

{\it \textbf{First objection}}: In addition to the exciton coupled to the $\vQ$ photon, the formation of a trion requires a free electron (see Fig.~1(b)). Just from momentum conservation, we see that the re-emitted photon can have a momentum different from $\vQ$, even if the trion has a lifetime long enough to keep its  $\vQ+\vk$ momentum. The resulting different photon states ($\vQ+\vq$) lead to an \textit{intrinsic} broadening of the emitted photon energy very similar to the \textit{extrinsic} broadening that prevents the formation of exciton-polariton in the case of  the aforementioned weak coupling. To form a trion-polariton, additional physics is required to prevent photon absorption by forcing the photon back to its initial momentum.

\begin{figure}[t!]
\centering
  \includegraphics[trim=1.3cm 3.1cm 1.3cm 3cm,clip,width=3.4in] {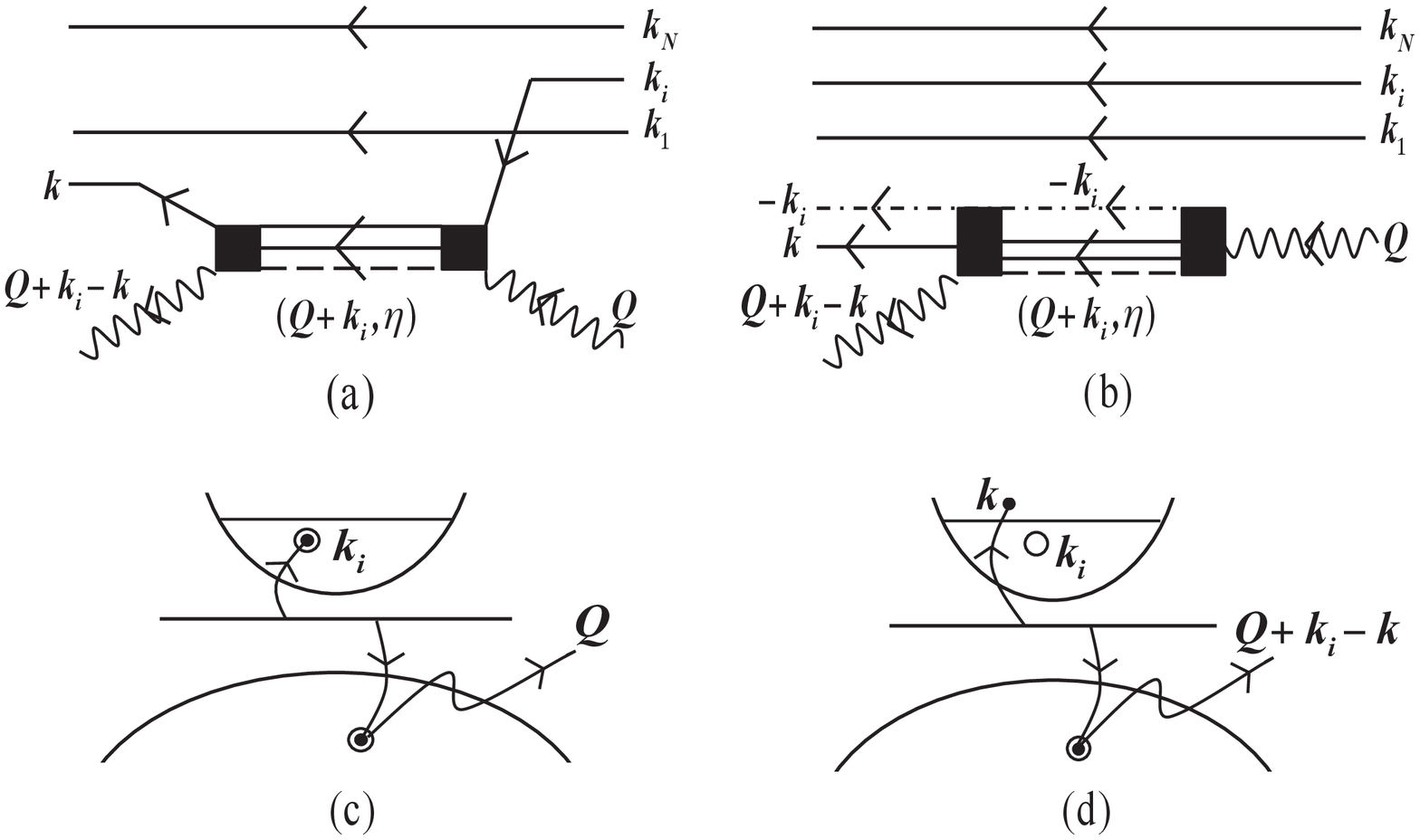}
 \vspace{-0.1cm}
\caption{\small (a) A $\vQ$ photon and a $\vk_i$ electron taken from the conduction Fermi sea couple to trions $(\vK,\eta)$ with $\vK=\vQ+\vk_i$. Such trion recombines into a photon-electron pair $(\vQ+\vk_i-\vk,\vk)$ with $\vk=\vk_i$ or $|\vk|>|\vk_N|=k_F$ due to Pauli blocking with the Fermi sea. (b) Equivalently, a $\vQ$ photon in the presence of a Fermi sea transforms into a $(\vQ+\vk_i,\eta)$ trion and a $(-\vk_i)$ conduction hole. The trion--conduction-hole pair can either transform back into the same $\vQ$ photon with the $\vk$ electron refilling the $(-\vk_i)$ hole, as in (c), or into a photon having a far larger energy than the $\vQ$ photon if the released electron goes above the Fermi sea, as in (d). }
\label{fig:2}
\end{figure}

 This physics is Pauli blocking with the electrons of a conduction-band Fermi sea. When a $\vk_i$ electron is taken from the Fermi sea to form a trion, it leaves a conduction-hole (see Figs.~2(a,b)). Due to Pauli blocking with the other Fermi-sea electrons, the $\vk$ electron that is released after trion recombination can have a momentum equal to $\vk_i$, thereby refilling the conduction-hole (see Fig.~2(c)), or larger than the Fermi momentum $k_F$ (see Fig.~2(d)). As photons in the visible spectrum have momenta far smaller than typical electron momenta, a $(\vQ+\vk)$ photon with $|\vk|>k_F$ would have an energy much higher than the $\vQ$-photon energy, bringing it out of trion resonance. So, the dominant process corresponds to $\vk=\vk_i$. We then have a  photon--trion-hole chain (see Fig.~1(c)) similar to the photon-exciton chain leading to exciton-polariton, the photon momentum being pinned to $\vQ$ by Pauli blocking with the Fermi sea.

{\it\textbf{ Second objection}}: Overcoming the above objection is not enough to produce a trion-polariton because
the coupling between a trion and a photon plus a $\vk$ electron is vanishingly small: two plane waves, one for the photon and one for the electron, then transform into the plane wave of the bound-trion center of mass. The formation of a bound trion requires the localization of an electron, originally delocalized over the sample volume $L^D$, into the trion relative-motion volume $a_T^D$. This brings a dramatic reduction factor $(a_T/L)^D$ to the photon-trion coupling compared to photon-exciton value \cite{poorCoupling,Shiauprb2012,T1}. \

As shown more in detail below, the Fermi sea overcomes this vanishingly small coupling by providing a volume-linear trion-hole subspace to which the $\vQ$ photon is coherently coupled, and which can render the photon-trion effective coupling of the order of the photon-exciton coupling.  \

{\it \textbf{Third objection}}: Because electrons are fermions, the Fermi sea produces an energy broadening of the photon resonance to trion, which occurs at
\be
\omega_\vQ +\frac{\vk_i^2}{2m_e}\simeq E_{gap}+ \mathcal{E}^{(T)}_{\vQ+\vk_i; \eta_0}
        \label{omegaQ}
\ee
if we forget interaction between the trion and the Fermi-sea electrons. $E_{gap}$ is the band gap, and $ E_{gap}+\mathcal{E}^{(T)}_{\vQ+\vk_i;\eta_0}$ with $\mathcal{E}^{(T)}_{\vQ+\vk_i;\eta_0}=\mathcal{E}^{(T)}_{\eta_0} +(\vQ+\vk_i)^2/2(2m_e+m_h)$ is the energy of a trion with  center-of-mass momentum $(\vQ+\vk_i)$ and  relative-motion ground state index $\eta_0$.
So, for typical photon momentum $|\vQ|\ll|\vk_i|\leq k_F$, the trion-polariton energy $\omega_\vQ$ has an energy broadening that varies from 1 to 2/3 Fermi energy, depending on the electron-hole mass ratio $m_e/m_h$. Many-body interactions with Fermi-sea electrons further produce a sharp low-energy side to the trion resonance due to low-energy excitations close to the Fermi level, and a long high-energy tail\cite{C5}. \

Furthermore, when a large number of trions are simultaneously present due to the absorption of many  photons, these trions, which are fermions, have different energies; this constitutes another source of trion-resonance broadening.

{\it \textbf{Fourth objection}}: 
In the absence of Fermi sea, the momenta $\vk$ of the electron attracted by the exciton to form a bound trion are predominantly smaller than $1/a_T$, the trion Bohr radius $a_T$ being a few times larger than the exciton Bohr radius $a_X$. In the presence of a Fermi sea, the $\vk$ states with $0\leqslant|\vk |\leqslant k_F$ are Pauli-blocked. This blocking decreases the trion binding energy. To keep a sizable binding, the Fermi momentum $k_F$ has to be smaller than $1/a_T$. Coulomb screening by Fermi-sea electrons also tends to reduce the trion binding.

\begin{figure}[t!]
\centering
\includegraphics[trim=0cm 0cm 0cm 0cm,clip,width=3.3in]{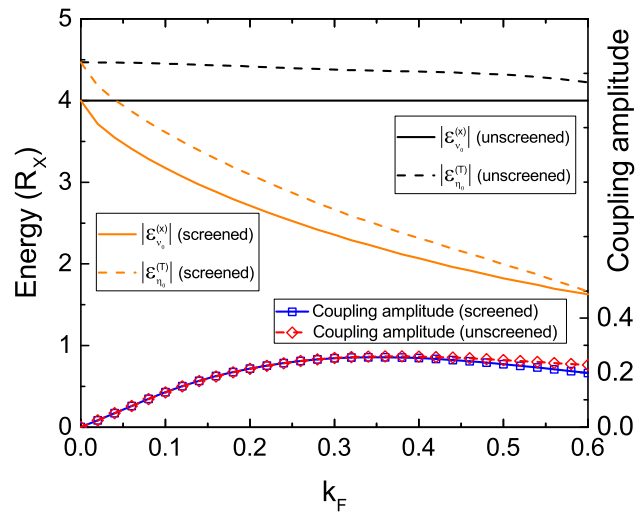}
   \caption{\small(color online) Upper curves: 2D energies of ground-state exciton $|\mathcal{E}^{(X)}_{\nu_0}(k_F)|$ and ground-state trion $|\mathcal{E}^{(T)}_{\eta_0}(k_F)|$ (in 3D Rydberg unit $R_X$)
  , with and without Coulomb screening, as a function of the Fermi momentum $k_F$ (in $1/a_X$ unit) for infinite valence-hole mass and a Fermi sea polarization opposite to that of the electron making the exciton. 
  Lower curves: ratio of the photon-trion coupling amplitude and photon-exciton coupling amplitude, and multiplied by $\sqrt{N}$, where $N$ is the number of Fermi-sea electrons.}
   \label{fig:3}
\end{figure}

 Figure \ref{fig:3} shows  the calculated two-dimensional (2D) energy of a ground-state exciton $|\mathcal{E}^{(X)}_{\nu_0}(k_F)|$ and a ground-state trion $|\mathcal{E}^{(T)}_{\eta_0}(k_F)|$, in the presence of a  spin-polarized Fermi sea having electronic spin opposite to the spin of the electron making the exciton (See Appendix for details). Screening by the 2D Fermi sea with Fermi momentum $k_F$ is included in a standard way, by replacing the 2D Coulomb potential $V_\vq=2\pi e^2/\epsilon_{sc}L^2q$ by
$V_\vq/\kappa(q)$ with $\kappa(q)=1+(2/qa_X)\Big[1-\Theta(q-2k_F)\sqrt{1-(2k_F/q)^2}\Big]$ where $\Theta(x)$ is the Heaviside function\cite{Stern}.\

 Because we have chosen to take Fermi sea electrons with a spin different from the spin of the electron making the exciton, Pauli blocking does not affect the exciton energy; so, in the absence of screening, $|\mathcal{E}^{(X)}_{\nu_0}(k_F)|$ stays equal to $4R_X$ where $R_X$ is the 3D exciton Rydberg. By contrast, Pauli blocking reduces the binding energy of the ground-state trion $|\mathcal{E}^{(T)}_{\eta_0}(k_F)|$, starting from $4.47R_X$ \cite{Stebe1989,Thilagam1997,Sergeev2001} in the absence of Fermi sea when the hole mass is infinite. The inclusion of Fermi-sea screening reduces even more the trion energy. It also affects the exciton energy, although the reduction is smaller. The two energies get very close for $k_Fa_X\sim 0.6$, making the Pauli-blocked electron of the trion essentially unbound: the trion is no longer a valid description of the system and the exciton has to be taken into consideration. So, in order to have a real trion well separated from the exciton, it is necessary to have a moderately dense Fermi sea with a  Fermi energy smaller than the trion binding, in agreement with qualitative arguments.  \

Figure 3 also shows the ratio of the calculated photon-trion coupling $\Omega_{\eta_0,0;\vQ=0,\vk_i}$ (see Eq.~(9)) for $|\vk_i|\rightarrow k_F$  and the photon-exciton coupling $\Omega_{\vQ=0;\nu_0}$, multiplied by $\sqrt{N}$ where $N=L^2k_F^2/4\pi$ is the number of electron states in the 2D Fermi sea. Its value increases with Fermi energy as expected, with a very small screening effect. The photon-trion coupling $\Omega_{\eta_0,0;\vQ=0,\vk_i}$ is insensitive to $\vk_i$ if Fermi momentum $k_F$ is not too large.

 \textit{\textbf{All this leads to an optimal Fermi sea}} for trion-polariton formation not too large in order for the trion subspace coupled to a $\vQ$ photon to be quasi-degenerate while enhancing the reduction factor up to $N(a_T/L)^D$ of the order of $1$ for the photon-trion coupling to be of the order of the exciton-photon coupling. Since the electron density $N/L^D$ scales as $k_F^D$,  this corresponds to an electron Fermi energy $E_F$ of the order of a fraction of the trion binding. Note that a far smaller Fermi sea would be sufficient to pin the photon momentum to its original $\vQ$ value. Also note that the positively-charged conduction-hole can further form a bound state with the negatively-charged trion (see Fig.~2(b)); the trion-hole polariton then transforms into an excitonic dipole interacting with the dipole of a conduction electron-hole pair, or more generally, a cloud of such pairs. However, as the binding between trion and hole stays very small for $k_Fa_X\ll 1$, the trion-hole object is a more accurate description when the Fermi sea is dilute, than the exciton dressed by a cloud of conduction electron-hole pairs, as proposed in Refs.\cite{Sidler2016,Efimkin2016}.  \


{\it \textbf{Additional comments}:}
The above trion-polariton picture corresponds to a coupling between two composite fermions, namely a photon-electron pair and a trion. The same physics can be described in a more standard way in terms of cobosons: one just has to consider the Fermi sea as filled, and allows it to ``boil" with excitation of conduction electron-hole pairs. The $\vQ$ photon then transforms (see Fig.~1(c)) into a bosonic pair of $(\vQ+\vk_i,\eta)$ trion and $(-\vk_i)$ conduction-hole, the resulting trion-polariton being a quite complex coboson, with a promise of very interesting many-body physics.


  The Fermi sea with a $\vk_i$ state empty is surely shaken up by Coulomb interaction with the trion: this interaction can scatter the empty state from $\vk_i$ to $\vk_j\neq\vk_i$ or excite additional conduction electron-hole pairs. While the former process does not change the dimensionality $N$ of the photon-coupled subspace made of $\vQ+\vk_i$ trion and $(-\vk_i)$ conduction-hole, the recombination of a trion in the presence of conduction electron-hole pairs produces photons having an energy much higher than the $\vQ$ photon energy.  So, the inclusion of such Coulomb processes does not change our conclusions.


{\it \textbf{Formalism}.} We now outline how the above understanding follows from the coboson many-body formalism. The key hinges on the description of trion as an exciton interacting with an electron\cite{T1,T3,T4,T2,C8}.\

The first step is to write the trion on the exciton-electron basis. For two electrons located at $(\vr_{e},\vr_{e'})$ and one hole located at $\vr_h$, the relevant trion coordinates\cite{T2,T1} are the trion center-of-mass position $\vR_T$, the electron-hole distance  $\vr=\vr_e-\vr_h$ in the $(e,h)$ exciton and the distance $\vu=\vr_{e'}-(m_e\vr_e+m_h\vr_h)/(m_e+m_h)$ between the $e'$ electron and the exciton center of mass.  For opposite-spin electrons in singlet $(S=0)$ or triplet $(S=1)$ state and hole ``spin" $m=\pm 3/2$, the trion creation operator can be written in terms of electron and exciton creation operators, $a^\dag$ and $B^\dag$, as
\be
\hspace{-0.06cm}T^\dag_{\vK,\eta;S,m}=\sum_{\vp,\nu} a^\dag_{\vp+\beta_e \vK;\frac{1}{2}}B^\dag_{-\vp+\beta_X\vK,\nu;-\frac{1}{2},m} \lan \vp,\nu|\eta ,S\ran
\ee
where $\vK$ is the trion center-of-mass momentum and $\beta_e=1-\beta_X=m_e/(2m_e+m_h)$ in order for $\vp$ to be the relative-motion momentum. The trion relative-motion wave function $\lan \vu,\vr|\eta ,S\ran$
appears in the above equation through its Fourier transform
\be
\lan \vp,\nu|\eta ,S\ran=\int \int d\vu \, d\vr\,\lan \vp|\vu\ran \lan \nu|\vr\ran \lan \vu,\vr|\eta ,S\ran\, .
\ee
Conversely, the creation operator for a electron-exciton-pair can be written in terms of trion operators as
\be
a^\dag_{\vp+\beta_e\vK;\frac{1}{2}}B^\dag_{-\vp+\beta_X\vK,\nu;-\frac{1}{2},m}=\sum_{\eta,S=(0,1)}T^\dag_{\vK,\eta;S,m}\lan \eta,S|\nu,\vp\ran\, .
\ee

The photon-semiconductor Hamiltonian reads $H=H_{ph}+H_{sc}+W_{ph-sc}$ with $H_{sc}=H_e+H_h+V_{ee}+V_{eh}+V_{eh}$ while for circularly polarized photons $(\vQ;\sigma=1)$,
\be
W_{ph-sc}=\sum_{\vQ}\alpha_{\vQ;1} \sum_\nu \Omega_{\vQ\nu}B^\dag_{\vQ,\nu; -\frac{1}{2},\frac{3}{2}}+h.c.\, ,
\ee
the photon-exciton coupling\cite{T1} being proportional to the vacuum Rabi coupling $\Omega_\vQ$ and the exciton relative-motion wave function $\lan \nu|\vr\ran$, namely $\Omega_{\vQ\nu}=\Omega_\vQ L^{D/2}\lan \nu|\vr=0\ran $.\

To understand the assisting role played by the Fermi sea in the formation of trion-polariton, let us first consider a Fermi sea having a single electron $(\vk;1/2)$. To get the subspace spanned by the photon-electron pair, we use Eqs.~(4,5) to get, for $|v\ran$ denoting the vacuum state,
\bea
\lefteqn{\left(H-\omega_\vQ-\va^{(e)}_\vk\right)\alpha^\dag_{\vQ;1}a^\dag_{\vk;\frac{1}{2}}|v\ran=\sum_\nu \Omega_{\vQ\nu}B^\dag_{\vQ,\nu;-\frac{1}{2},\frac{3}{2}}a^\dag_{\vk;\frac{1}{2}}|v\ran}\hspace{8cm}\nn\\
=\sum_{\eta,S}\Omega_{\eta,S;\vQ,\vk}\,  T^\dag_{\vk+\vQ,\eta;S,\frac{3}{2}}|v\ran\, ,
\eea
 the coupling to trion\cite{T1} being given by $\Omega_{\eta,S;\vQ,\vk}=\Omega_\vQ L^{D/2}\lan \eta,S|\vr=0,\beta_X\vk-\beta_e\vQ\ran$. Using Eqs.~(2,5), we also get
\bea
\lefteqn{\left(H-E^{(T)}_{\vQ+\vk;\eta,S}\right)T^\dag_{\vk+\vQ,\eta;S,\frac{3}{2}}|v\ran}\hspace{1cm}\nn\\
&=&\sum_\vq \Omega^*_{\eta,S;\vQ+\vq,\vk-\vq}\, \alpha^\dag_{\vQ+\vq;1}a^\dag_{\vk-\vq;\frac{1}{2}}|v\ran\, .
\eea
The above equations readily show that the coupled subspace contains photons with momentum $\vQ+\vq$; they also show that the energy splitting at resonance with the ground-state trion $(\eta_0,S=0)$ is controlled by $|\Omega_{\eta_0,S=0;\vQ,\vk}|^2$ which is $(a_T/L)^D$ smaller than the photon-exciton splitting\cite{poorCoupling}, as previously explained. So, a trion-polariton with just one electron $\vk$ cannot be formed due to this too small coupling. \

Let us now consider the same $(\vQ,\sigma=1)$ photon but $N$ spin-(1/2) electrons in their ground state $(H_{sc}-\mathcal{E}_N)|F_{N,1/2}\ran=0$. Equation (5) then gives
\bea
(H{-}\omega_\vQ{-}\mathcal{E}_N)\alpha^\dag_{\vQ,1}|F_{N,\frac{1}{2}}\ran{=}\sum_\nu \Omega_{\vQ\nu}B^\dag_{\vQ,\nu;{-}\frac{1}{2},\frac{3}{2}}|F_{N,\frac{1}{2}}\ran.
\eea
To force the trion into the problem, we single out a $\vk_i$ electron from the Fermi sea by noting that, since $[a_{\vk;1/2},a^\dag_{\vk;1/2}]_+=1$, we can rewrite $|F_{N,1/2}\ran$ as $ a^\dag_{\vk_i;1/2}a_{\vk_i;1/2}|F_{N,1/2}\ran$ for any $\vk_i$ in the Fermi sea. Equation (4) then gives the RHS of the above equation in terms of trions as
\be
\sum_{\eta,S=(0,1)}\Omega_{\eta,S;\vQ,\vk_i} \,  T^\dag_{\vQ+\vk_i,\eta;S,\frac{3}{2}} a_{\vk_i;\frac{1}{2}}|F_{N,\frac{1}{2}}\ran\, .
\ee
Equation (5) also gives, by neglecting Coulomb interaction between the trion and the Fermi sea with the $\vk_i$ state empty for the reasons previously discussed, 
\bea
\left(H-E^{(T)}_{\vQ+\vk_i;\eta,S}
+\va_{\vk_i}^{(e)}
-\mathcal{E}_N\right)
T^\dag_{\vk_i+\vQ,\eta;S,\frac{3}{2}}
 a_{\vk_i;\frac{1}{2}}
 |F_{N,\frac{1}{2}}\ran\nn\\
\simeq\sum_\vq \Omega^*_{\eta,S;\vQ+\vq,\vk_i-\vq}\, \alpha^\dag_{\vQ+\vq;1}a^\dag_{\vk_i-\vq;\frac{1}{2}}a_{\vk_i;\frac{1}{2}}|F_{N,\frac{1}{2}}\ran\, \,\,\,.
\eea
This equation reduces to Eq.~(7) when the Fermi sea contains one electron only. When it contains a large number of electrons, Pauli blocking forces the electron momentum $\vk_i-\vq$ in the above sum to be either equal to $\vk_i$ or above the Fermi level---the latter being excluded because too far from the initial energy.
This reduces the $\vq$ sum to its $\vq=0$ term: the photon then keeps its initial momentum $\vQ$, in the same way as for exciton-polariton with strong coupling; however, this condition is here fulfilled  with the help of Pauli blocking with the Fermi sea.\

Still, the coupling $|\Omega_{\eta_0,0;\vQ,\vk_i}|^2$ between the photon $\vQ$ and the trion--conduction-hole pair $(\vQ+\vk_i,-\vk_i)$  is $(a_T/L)^D$ smaller than the one for exciton-polariton. A volume-linear factor is required to overcome this reduction factor. To get it, we
 note that the $\vQ$ photon has similar vanishingly-small couplings to the $N$ trion-hole pairs made with any of the $N$ Fermi sea electrons $\vk_j$. Diagonalization within this coupled subspace transforms 
\be
 \hspace{-0.14cm}(\omega_\vQ{+}\mathcal{E}_N{-}\mathbb{E}) \big(E^{(T)}_{\vQ{+}\vk;\eta_0,0}{-}\va_{\vk}^{(e)}{+}\mathcal{E}_N{-}\mathbb{E}\big)
{=}
|\Omega_{\eta_0,0;\vQ,\vk}|^2
\ee
that gives the system energy  when the Fermi sea contains a single electron $\vk$, into
 \bea
(\omega_\vQ+\mathcal{E}_N-\mathbb{E})\prod_{i=1}^N\left(E^{(T)}_{\vQ+\vk_i;\eta_0,0}-\va_{\vk_i}^{(e)}+\mathcal{E}_N-\mathbb{E}\right)\hspace{0.8cm}\\
=\sum_{i=1}^N|\Omega_{\eta_0,0;\vQ,\vk_i}|^2\prod_{j=1\atop j\neq i}^N\left(E^{(T)}_{\vQ+\vk_j;\eta_0,0}-\va_{\vk_j}^{(e)}+\mathcal{E}_N-\mathbb{E}\right)\, .\nn
 \eea
By taking the trion--conduction-hole states as quasi-degenerate, Eq.~(12) then reduces to Eq.~(11) with $|\Omega_{\eta_0,0;\vQ,\vk}|^2$ multiplied by $N$. This leads to a trion-polariton splitting as large as the exciton-polariton splitting for $N(a_T/L)^D\sim1$. \


  In the above calculations, 
  we neglected Coulomb processes between trions and Fermi sea electrons; these processes either change $\vk_i$ into $\vk_j$, or create additional conduction electron-hole pairs, but they do not qualitatively change the above results, as previously explained. We also took the Fermi sea as fully polarized to elucidate the physics of this novel polariton in the most simple way. Adding spin-$(-1/2)$ electrons reduces the exciton binding due to Pauli blocking with same-spin electrons, but mostly broadens the trion resonance. Consequences raised by these issues, including the Fermi edge singularity\cite{CN1,Mahan,MCT1,Baeten2014} associated with the sudden appearance of a trion in the presence of a dense Fermi sea, will be studied elsewhere.

  Strong coupling regime in the presence of an unpolarized conduction Fermi sea has been recently investigated using transition metal dichalcogenide (MoSe$_2$) monolayers embedded in microcavity\cite{Sidler2016}. The novel state found below the exciton-polariton  has been partially explained in terms of ``Fermi polaron-polariton" using a model based on a {\it rigid} exciton interacting with photon and electrons. Moreover, Coulomb screening and mostly Pauli blocking induced by Fermi electrons on the trion and the exciton have been neglected or included in a crude way. A more sophisticated model\cite{Efimkin2016} but along the same line has been proposed to study  optical absorption. These models do miss the rich fermion-exchange physics associated with the trion structure, which is crucial, and often dominant, in problems dealing with composite bosons, like excitons, and composite fermions, like trions. These  recent experiments have been unfortunately performed in a quite complex configuration; so, it is rather difficult to make a direct comparison with the present study which mostly focuses on the fundamental aspects of photon and trion in the presence of a Fermi sea.

\textbf{Appendix on numerical calculations}\

To obtain the trion ground state in the presence of  a frozen Fermi sea $|F_{N,1/2}\ran$,  we  for simplicity consider the valence-hole with an infinite mass, hence as fixed. We then expand the $(\vk ,-1/2)$ and $(\vk' ,1/2)$ electronic states of the trion  as
$\sum_\vk
 \sum_{|\vk'|> k_F}
 G_{\vk,\vk'} \,\,
 a^\dagger_{\vk;-\frac{1}{2}}a^\dagger_{\vk';\frac{1}{2}} $, with $\vk'$ outside the Fermi sea due to Pauli blocking it induces. \
 
To minimize the Hamiltonian mean value, we have expanded the trial function $G_{\vk,\vk'}$ on products of 2D cylindrical functions which, for $\vr=(r,\varphi)$ and $\ell=(0,\pm1,\pm2\cdots)$, read as $\Phi_{\ell,\lambda}(\vr)=e^{i\ell\varphi }e^{-\lambda r} r^ {1-\delta_{\ell,0}}$, or better its Fourier transform $\Phi_{\ell,\lambda;\vk}$. The 2D exciton ground state corresponds to $(\ell=0,\lambda=2)$ for $r$ in 3D Bohr radius $a_X$. So, 
$G_{\vk,\vk'}=
\sum_{\ell,\lambda} \,\,
 \sum_{\ell',\lambda'}
 g_{\ell,\lambda;\ell',\lambda'}\,\,
 \Phi_{\ell,\lambda;\vk}\,\,
 \Phi_{\ell',\lambda';\vk'}$.
In practice, we have restricted $\ell$ to $(0,\pm1,\pm2)$ and taken $\ell'=-\ell$ since the ground state has zero angular momentum. We have also used an even-tempered set of seven $(\lambda,\lambda')$ parameters. We then used the calculated trion ground-state wave function to obtain the ratio of the photon-trion coupling $\Omega_{\eta_0,0;\vQ=0,\vk}$ for $|\vk|\rightarrow k_F$ to the photon-exciton coupling $\Omega_{\vQ=0;\nu_0}$, shown in Fig.~3.

\textbf{Acknowledgments}: This work has been initiated by discussions with A. Imamoglu and G. Morigi, during the program  ``Many-Body Physics with Light (2015)" that M. C. attended at the Kavli Institute for Theoretical Physics, University of California at Santa Barbara. This work was supported in part by the Ministry of Science and Technology, Taiwan under Contract No. MOST 104-2112-M-001-009-MY2.


\begin{thebibliography}{99}


\bibitem{polCond1} H. Deng, G. Weihs, C. Santori, J. Bloch, and Y. Yamamoto, Science {\bf298}, 199 (2002).
\bibitem{polCond2} J. Kasprzak, M. Richard, S. Kundermann, A. Baas, P. Jeambrun, J. M. J. Keeling, F. M. Marchetti, M. H. Szyma\'{n}ska, R. Andr\'{e}, J. L. Staehli, V. Savona, P. B. Littlewood, B. Deveaud, and L. S. Dang, Nature {\bf443}, 409 (2006).
\bibitem{polCond3} H. Deng, D. Press, S. G\"{o}tzinger, G. S. Solomon., R. Hey, K. H. Ploog, and Y. Yamamoto, Phys. Rev. Lett. {\bf97}, 146402 (2006).
\bibitem{polCond4} R. Balili, V. Hartwell, D. Snoke, L. Pfeiffer, and K. West, Science {\bf316}, 1007 (2007).

\bibitem{polPredict}  J. J. Hopfield, Phys. Rev. {\bf112}, 1555 (1958).

\bibitem{polSeen}  C. Weisbuch, M. Nishioka, A. Ishikawa, and Y. Arakawa, Phys. Rev. Lett. {\bf69}, 3314 (1992).

\bibitem{DubinEPL} F. Dubin, M. Combescot, and B. Roulet, Europhys. Lett. {\bf69}, 931 (2005).























\bibitem{trionPredict} M. A. Lampert, Phys. Rev. Lett. {\bf1}, 450 (1958).

\bibitem{trionseen1} S. Narita and M. Taniguchi, Phys. Rev. Lett. {\bf36}, 913 (1976).
\bibitem{trionseen2} M. Taniguchi and S. Narita, Solid State Commun. {\bf20}, 131 (1976).
\bibitem{trionseen3} K. Kheng, R. T. Cox, Y. Merle d'Aubign\'{e}, F. Bassani, K. Saminadayar, and S. Tatarenko, Phys. Rev. Lett. {\bf71}, 1752 (1993).
\bibitem{trionseen4} G. Finkelstein, H. Shtrikman, and I. Bar-Joseph, Phys. Rev. Lett. {\bf74}, 976 (1995).
\bibitem{trionseen5} A. J. Shields, J. L. Osborne, M. Y. Simmons, M. Pepper, and D. A. Ritchie, Phys. Rev. B {\bf52}, R5523 (1995).
\bibitem{trionseen6} H. Buhmann, L. Mansouri, J.Wang, P. H. Beton, N. Mori, L. Eaves, M. Henini, and M. Potemski, Phys. Rev. B {\bf51}, 7969 (1995).


\bibitem{Sidler2016} M. Sidler, P. Back, O. Cotlet, A. Srivastava, T. Fink, M. Kroner, E. Demler, and A. Imamoglu, Nat. Phys. (2016) doi:10.1038/nphys3949.


\bibitem{Jones2013}A. M. Jones,	 H. Yu,	 N. J. Ghimire,	 S. Wu,	 G. Aivazian,	 J. S. Ross, B. Zhao,	 J. Yan,	 D. G. Mandrus,	 D. Xiao,	 W. Yao, and X. Xu, Nature Nanotech. {\bf8}, 634 (2013).

\bibitem{Ross2013} J. S. Ross, S. Wu, H. Yu, N. J. Ghimire, A. M. Jones, G. Aivazian, J. Yan, D. G. Mandrus, D. Xiao, W. Yao, and X. Xu, Nature Commun. {\bf4}, 1474 (2013).

\bibitem{Mak2013} K. F. Mak,	 K. He,	 C. Lee,	 G. H. Lee,	 J. Hone,	 T. F. Heinz, and J. Shan, Nature Mater. {\bf12}, 207 (2013).
\bibitem{Xu2014} X. Xu,	 W. Yao,	 D. Xiao, and T. F. Heinz, Nature Phys. {\bf10}, 343 (2014).

\bibitem{poorCoupling} M. Combescot and J. Tribollet, Solid State Comm. {\bf128}, 273 (2003).
\bibitem{Shiauprb2012}S.Y. Shiau, M. Combescot, and Y.C. Chang, Phys. Rev. B {\bf86}, 115210 (2012).


\bibitem{T1} M. Combescot and S.-Y. Shiau, {\it Excitons and Cooper Pairs}, Oxford Univ. Press, Oxford (2015).

\bibitem{C5} M. Combescot, J. Tribollet, G. karczewski, F. Bernardot, C. Testelin, and M. Chamarro, Europhys. Lett. {\bf71}, 431 (2005).

\bibitem{Stern} F. Stern, Phys. Rev. lett. {\bf18}, 546 (1967).


\bibitem{Stebe1989} B. St\'{e}b\'{e} and A. Ainane, Superlattices and Microstructures {\bf 5}, 545 (1989).

\bibitem{Thilagam1997} A. Thilagam, Phys. Rev. B {\bf 55}, 7804 (1997).
 \bibitem{Sergeev2001} R. A. Sergeev and R. A. Suris, Nanotechnology {\bf 12}, 597 (2001).


\bibitem{Efimkin2016} D. K. Efimkin and A. H. MacDonald, arXiv: 1609.06329.



\bibitem{T2} M. Combescot, Euro. Phys. J. B {\bf33}, 311 (2003)

\bibitem{T3} M. Combescot, O. Betbeder-Matibet, Solid State Com. {\bf126}, 687 (2003).

\bibitem{T4} M. Combescot, O. Betbeder-Matibet, and F. Dubin, Euro. Phys. J. B {\bf42}, 63 (2004).

\bibitem{C8} M. Combescot and O. Betbeder-Matibet,
Phys. Rev. Lett. {\bf104}, 206404 (2010).

\bibitem{CN1} M. Combescot and P. Nozi\`{e}res, Journal de Physique {\bf32}, 913 (1971).

\bibitem{Mahan} G. D. Mahan, {\it Many-Particle Physics}, 2nd ed. (Plenum, New York, 1990).

\bibitem{MCT1} M. Combescot and C. Tanguy, Phys. Rev. B {\bf50}, 11484 (1994); C. Tanguy, and M. Combescot, Phys. Rev. B {\bf50}, 11499 (1994); Phys. Rev. B {\bf52}, 11698 (1995).

\bibitem{Baeten2014} M. Baeten and M. Wouters, Phys. Rev. B {\bf89}, 245301 (2014); Phys. Rev. B {\bf91}, 115313 (2015).



\end{thebibliography}
\end{document}